\documentclass[prx,twocolumn,,preprintnumbers,amsmath,amssymb]{revtex4}

\usepackage{graphicx,color}
\usepackage{dcolumn}
\usepackage{hyperref}
\usepackage{bm}
\usepackage{stmaryrd}
\usepackage{latexsym}
\usepackage{amssymb}
\usepackage{amsfonts}
\usepackage{amsmath}
\usepackage{epstopdf}

\begin{document}

\title{Evidence for unconventional superconductivity in a spinel oxide}

\author{Huanyi Xue,$^{1,*}$ Lijie Wang,$^{1,*}$ Zhongjie Wang,$^{1,*}$ Guanqun Zhang,$^{1}$ Wei Peng,$^{2,3}$ Shiwei Wu,$^{1,4}$ Chunlei Gao,$^{1,4,\dag}$ Zhenghua An,$^{1,4,\ddag}$ Yan Chen,$^{1}$ and Wei Li$^{1,\S}$}

\affiliation
{$^1$State Key Laboratory of Surface Physics and Department of Physics, Fudan University, Shanghai 200433, China\\
 $^2$State Key Laboratory of Functional Materials for Informatics, Shanghai Institute of Microsystem and Information Technology, and Center for Excellence in Superconducting Electronics, Chinese Academy of Sciences, Shanghai 200050, China\\
 $^3$Center of Materials Science and Optoelectronics Engineering, University of Chinese Academy of Sciences, Beijing 100049, China\\
 $^4$Institute for Nanoelectronic Devices and Quantum Computing, Fudan University, Shanghai 200433, China
 }

\date{\today}

\begin{abstract}
The charge frustration with the mixed-valence state inherent to LiTi$_2$O$_4$, which is found to be a unique spinel oxide superconductor, is the impetus for paying special attention to reveal the existence of intriguing superconducting properties. Here, we report a pronounced fourfold rotational symmetry of the superconductivity in high-quality single-crystalline LiTi$_2$O$_4$ (001) thin films. Both the magnetoresistivity and upper critical field under an applied magnetic field manifest striking fourfold oscillations deep inside the superconducting state, whereas the anisotropy vanishes in the normal state, demonstrating that it is an intrinsic property of the superconducting phase. We attribute this behavior to the unconventional $d$-wave superconducting Cooper pairs with the irreducible representation of $E_g$ protected by $O_h$ point group in LiTi$_2$O$_4$. Our findings demonstrate the unconventional character of the pairing interaction in a three-dimensional spinel oxide superconductor and shed new light on the pairing mechanism of unconventional superconductivity.
\end{abstract}

\maketitle

\section{Introduction}

The appearance of intriguing superconductivity arises from Cooper pairing between conducting electrons, and one of the prominent issues about superconductivity is its pairing symmetry, which provides a fundamental understanding of the Cooper pair formations in superconductivity~\cite{Ref1,Ref2,Ref3,Ref4}. In conventional superconductors, a condensate of Cooper pairs exhibits an isotropic $s$-wave pairing symmetry that is independent of directions over the entire Fermi surface~\cite{Ref5}. In unconventional superconductors, the superconducting Cooper pairs have anisotropic gap functions belonging to nontrivial irreducible representations of the crystal symmetry group that are not invariant under all symmetry elements~\cite{Ref1,Ref2}, such as the rotational symmetry breaking in layered two-dimensional copper oxide high-$T_c$ superconductors displaying anisotropic $d$-wave pairing~\cite{Ref6,Ref7,Ref8}. Notably, this rotational symmetry is spontaneously broken by strong Coulomb repulsion in the strongly correlated electron system of cuprate, which could lead to a novel many-body effect and give rise to Cooper pair states with orbital wave function with angular momentum greater than zero ($s$-wave)~\cite{DHLee2013}.

\begin{figure*}[t!]
\centering
\includegraphics[bb=90 60 500 410,width=11cm,height=9.5cm]{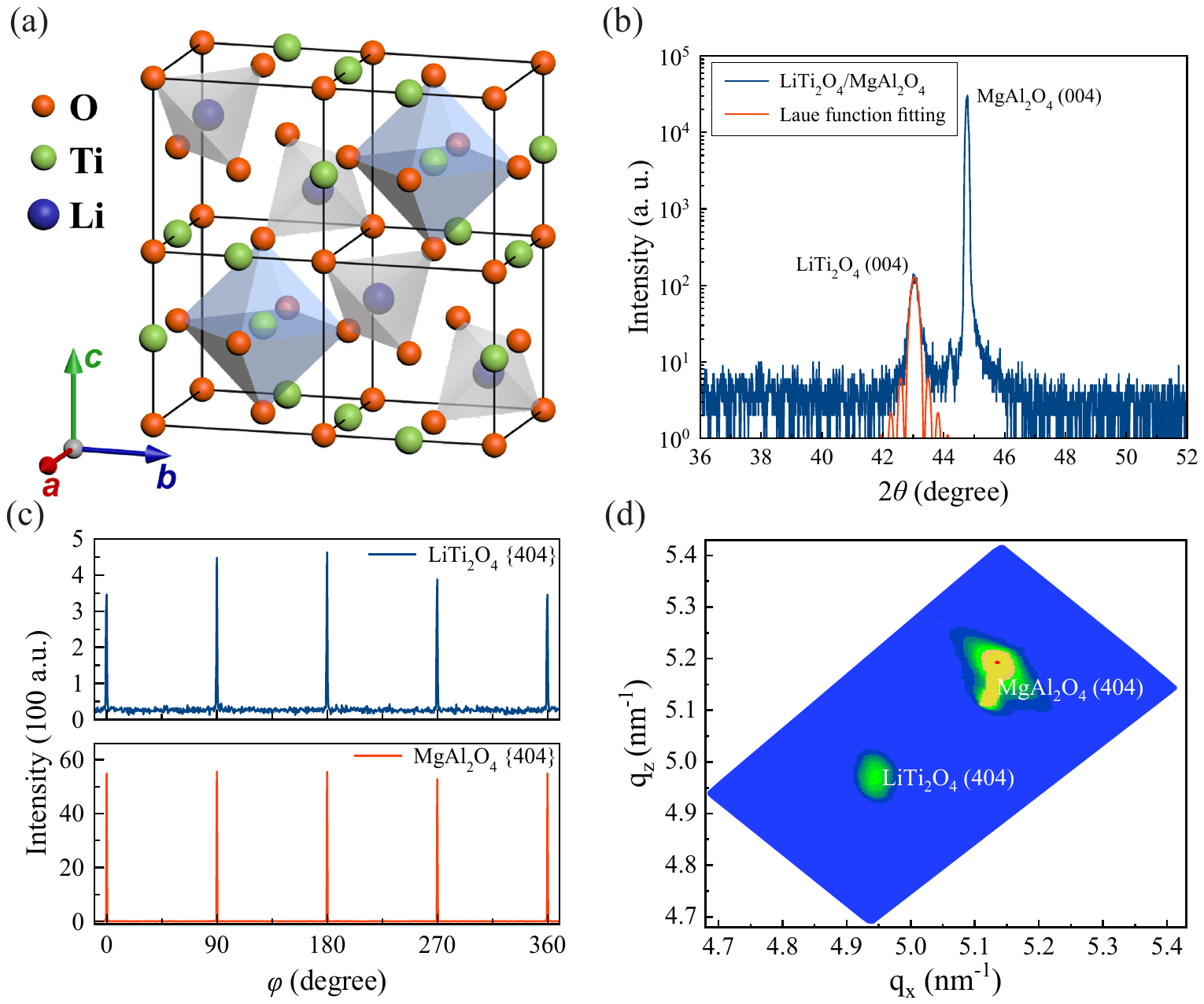}
\caption{Structural characteristics of LiTi$_2$O$_4$ thin films. (a) Schematic illustration of the crystal structure of spinel oxide LiTi$_2$O$_4$. (b) 2$\theta$ XRD spectrum of an epitaxial LiTi$_2$O$_4$ (001) thin film grown on a MgAl$_2$O$_4$ (001) substrate. The Laue function fitting is shown in (b). (c) $\varphi$-scan of the \{404\} diffraction planes for the LiTi$_2$O$_4$ thin film and MgAl$_2$O$_4$ substrate. Four peaks are uniformly distributed, displaying the in-plane fourfold rotational symmetry of the lattice, thus implying in-plane epitaxy. (d) Reciprocal space mapping of (404) peaks for both the thin film and substrate.}\label{fig1}
\end{figure*}

Spinel oxides are another striking class of strongly correlated electron systems that possess charge frustration with mixed valences and complicated interactions among charge, orbit, and spin induced by Jahn-Teller distortion, promoting many fascinating and appealing electronic phases. Among them, charge-frustrated lithium titanate (LiTi$_2$O$_4$) with mixed valences of Ti$^{3+}$: $3d^1$ and Ti$^{4+}$: $3d^0$ is a unique spinel oxide superconductor with a $T_c$ onset of 13 K~\cite{Ref9}. Previous specific heat measurements on polycrystalline LiTi$_2$O$_4$ suggest that this system is a candidate Bardeen-Cooper-Schrieffer (BCS)-like conventional $s$-wave superconductor~\cite{Ref10,SunCP2004}. However, non-negligible complex electron-electron correlations via spin fluctuations in superconductivity have been extensively revealed in X-ray absorption and resonant inelastic soft X-ray scattering~\cite{Ref11,Ref12}, nuclear magnetic resonance~\cite{Ref13}, and magnetic susceptibility measurements~\cite{Ref14}. Renewed measurements on high-quality LiTi$_2$O$_4$ thin films have revealed an anomalous magnetoresistivity in the normal state~\cite{Ref15} and a pseudogap opening at the Fermi energy~\cite{Ref16} through electrical transport measurement and scanning tunneling spectroscopy (STS), respectively, in recent years. These works allow us to conjecture that a possible unconventional and nontrivial superconductivity could be realized in LiTi$_2$O$_4$ driven by the intimate correlation between Cooper pairing and charge frustration associated with strong spin fluctuations~\cite{DHLee2013}.

In this paper, we revisit and discuss the possible pairing symmetry of the superconductivity in LiTi$_2$O$_4$ (001) thin films by using both the angular-resolved magnetoresistivity and upper critical field and find a pronounced fourfold rotational symmetry manifested deep inside the superconducting state that vanishes in the normal state. These results significantly demonstrate that the anisotropy with fourfold rotational symmetry is an intrinsic property of the superconducting phase in LiTi$_2$O$_4$, and thus, we classify the three-dimensional LiTi$_2$O$_4$ as a $d$-wave pairing unconventional superconductor with the irreducible representation of $E_g$ protected by $O_h$ point group. 

\section{Experimental results}

High-quality single-crystalline thin films of spinel LiTi$_2$O$_4$ (001) are epitaxially grown on MgAl$_2$O$_4$ (001) ($a_{\mathrm{MAO}}$ = 0.8080 nm) substrates by pulsed laser deposition. Bulk LiTi$_2$O$_4$ is a face-centered-cubic spinel structure at room-temperature with lattice parameter $a_{\mathrm{LTO}}$ = 0.8405 nm~\cite{Ref9} [Fig.~\ref{fig1}(a)], consisting of tetrahedral and octahedral sites occupied by lithium and titanium cations, respectively. The X-ray diffraction (XRD) $\theta$-$2\theta$ scans for the LiTi$_2$O$_4$ thin films are measured [Fig.~\ref{fig1}(b)], which display clear (004) Bragg reflection peaks of the films and substrates with the absence of additional peaks down to the sensitivity limit of the diffractometer, suggesting $c$-oriented growth of the LiTi$_2$O$_4$ thin films. The out-of-plane lattice parameter and thickness of the LiTi$_2$O$_4$ thin films are estimated to be 0.84 nm and 30.2 nm, respectively, deduced from the formula of Laue oscillation, in good agreement with that of the bulk~\cite{Ref9}. The $\varphi$-scan of the \{404\} diffraction planes displays fourfold rotational symmetry of the crystal lattice at the same angular positions [Fig.~\ref{fig1}(c)], implying in-plane ordering for the thin film and substrate. As shown in Fig.~\ref{fig1}(d), reciprocal space mapping of the (404) peak is carried out to further confirm epitaxial growth. These measurements, including large-area atomic force microscopy (see Fig. S1 and Sec. I in the Supplementary Material~\cite{SM}), demonstrate high-quality epitaxial growth of the LiTi$_2$O$_4$ films on MgAl$_2$O$_4$ substrates.

\begin{figure*}
\centering
\includegraphics[bb=90 60 490 410,width=11cm,height=9.5cm]{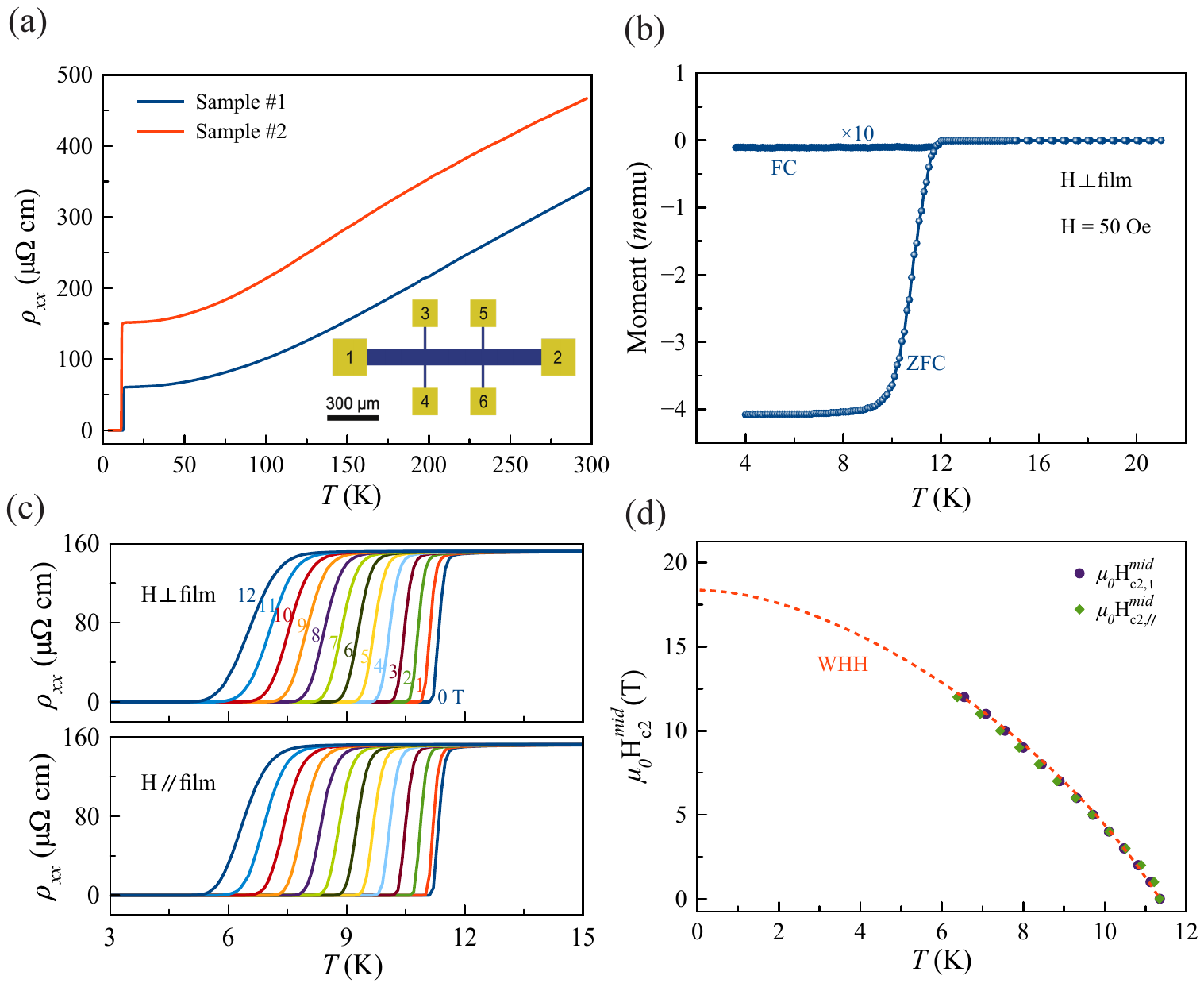}
\caption{Superconducting properties of LiTi$_2$O$_4$ (001) thin films. (a) Longitudinal electrical resistivity ($\rho_{xx}$) as a function of temperature at zero magnetic field for two representative LiTi$_2$O$_4$ (001) thin films (Samples \#1 and \#2). The Hall bar structure is schematically illustrated in the inset of (a). (b) Temperature-dependent DC magnetization of a LiTi$_2$O$_4$ (001) thin film (Sample \#1) in FC and ZFC modes with an applied out-of-plane magnetic field of 50 Oe. We note that the FC values in (b) are scaled by a factor of 10. (c) Magnetoresistivity for fields perpendicular and parallel to the plane surface of Sample \#2. (d) Corresponding temperature dependence of the upper critical fields $\mu_0$H$_{c2}^{mid}$ (H$_{c2,\parallel}^{mid}$ for the in-plane field along the a/b-axis and H$_{c2,\perp}^{mid}$ for the out-of-plane field), which are determined at half the values of $\rho_{xx}$ in (c). The red dashed line is fitted by WHH theory.}
\label{fig2}
\end{figure*}

Figure~\ref{fig2}(a) displays the temperature-dependent longitudinal electrical resistivity $\rho_{xx}$ on two representative as-grown LiTi$_2$O$_4$ thin films (Samples \#1 and \#2) with the Hall bar structure, schematically illustrated in the inset of Fig.~\ref{fig2}(a). A typical metallic behavior in the normal state and a sharp superconducting transition at $T_c$ (onset of 13 K for Sample \#1 and 12 K for Sample \#2) are clearly observed, with a narrow and sharp transition width of less than 0.5 K. Temperature-dependent direct current (DC) magnetization measurements in both field-cooling (FC) and zero-field-cooling (ZFC) modes under an applied out-of-plane magnetic field of 50 Oe are also carried out to further examine the superconductivity in a LiTi$_2$O$_4$ thin film (Sample \#1), as shown in Fig.~\ref{fig2}(b). The observed negative magnetic susceptibility, which is the ratio of the measured magnetization to the applied magnetic field, indicates the diamagnetism induced by the Meissner effect, unambiguously confirming the appearance of superconductivity below $T_c$. These results are highly reproducible and reasonably agree with previous electrical transport studies~\cite{Ref9,Ref15}. Furthermore, the magnetoresistivity $\rho_{xx}(\mu_0$H) (here, $\mu_0$ is the vacuum permeability) with fields perpendicular ($\mu_0$H$_{\perp}$) and parallel ($\mu_0$H$_{\parallel}$) to the sample plane surface of a LiTi$_2$O$_4$ thin film (Sample \#2) are shown in Fig.~\ref{fig2}(c) (Sample \#1 in Fig. S3 of the Supplemental Material~\cite{SM}). The fundamental superconducting behavior is clearly observed in that the superconducting critical fields $\mu_0$H$_{c2}^{mid}$ (H$_{c2,\parallel}^{mid}$ for the parallel field along the a/b-axis and H$_{c2,\perp}^{mid}$ for the perpendicular field) parallelly shift to a lower value, where the $\mu_0$H$_{c2}^{mid}$ values are evaluated at the midpoints of the normal-state resistivity. This shift arises from the magnetic field-induced orbital effect, which leads to the appearance of Abrikosov vortices and the formation of a regular array of vortex lines parallel to the magnetic field. As a result, the kinetic energy of superconducting currents around the vortex cores reduces the superconducting condensation energy~\cite{Ref17,Ref18}. Notably, neither direction of the magnetic field significantly affects the $\mu_0$H$_{c2}^{mid}$, which suggests that LiTi$_2$O$_4$ is a three-dimensional superconductor. For a quantitative estimation, the $\mu_0$H$_{c2}^{mid}$ are plotted as a function of temperature, and the H$_{c2}^{mid}$-$T$ phase diagram is shown in Fig.~\ref{fig2}(d). The Werthamer-Helfand-Hohenberg (WHH) model~\cite{Ref19} is further used to fit the $\mu_0$H$_{c2}^{mid}$ (also see Sec. II in the Supplemental Material~\cite{SM}). The extracted $\mu_0$H$_{c2}^{mid}$ at the zero temperature limit is 18.3 T, and the superconducting coherence length $\xi_{\mathrm{GL}}$($T$=0 K) is thus estimated from the Ginzburg-Landau formula~\cite{Ref17} $\mu_0$H$_{c2}^{mid}=\Phi_0/(2\pi\xi_{\mathrm{GL}}^2)$ with the fluxoid quantization $\Phi_0$ set as 4.2 nm, consistent with previous findings~\cite{Ref20,Ref21}.

\begin{figure*}[t!]
\centering
\includegraphics[bb=136 60 430 350,width=11cm,height=10cm]{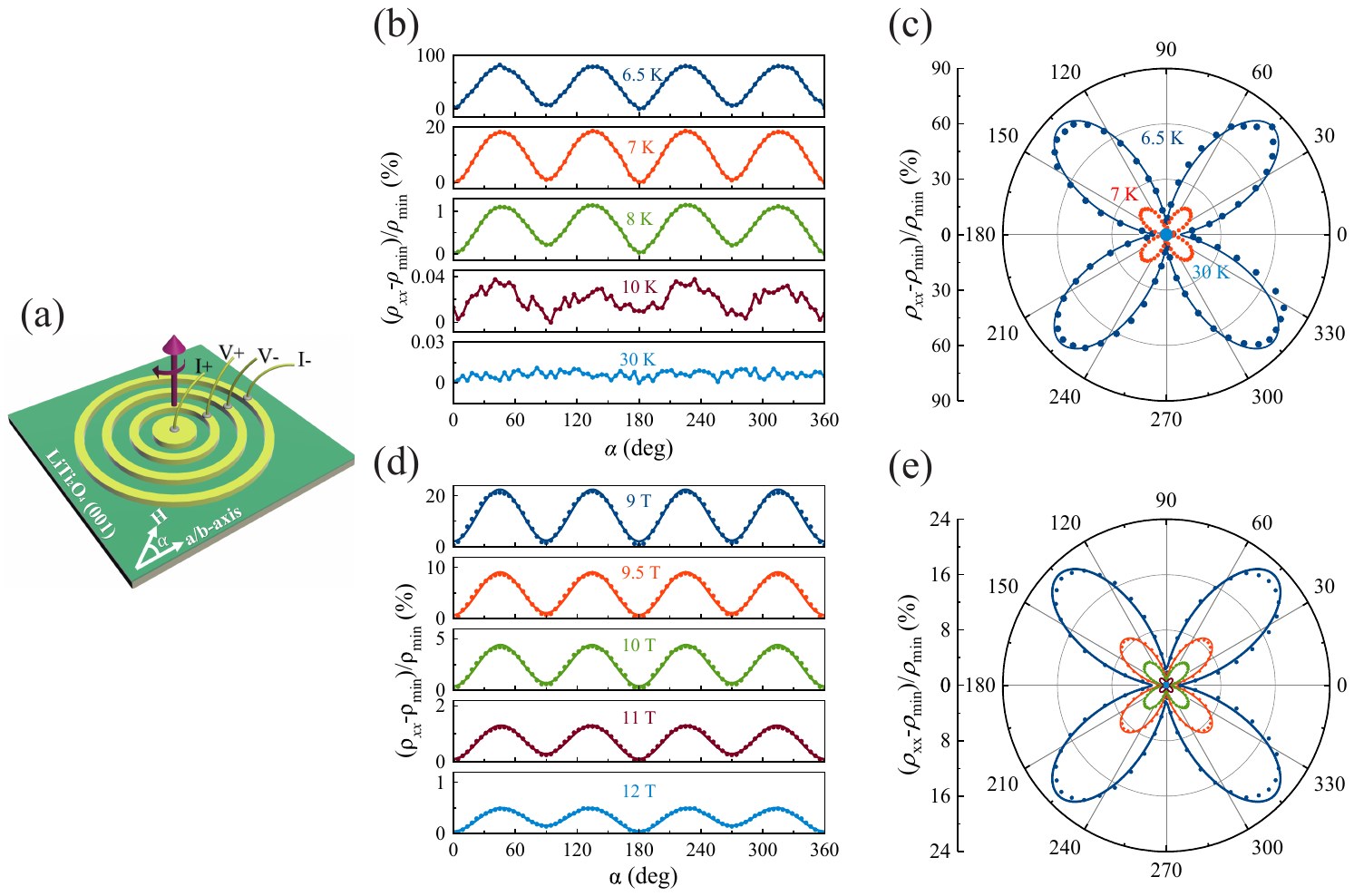}
\caption{In-plane magnetoresistivity signature of fourfold superconducting behavior in LiTi$_2$O$_4$ (001) thin films. (a) Schematic image of the Corbino-shaped device for in-plane angular-dependent magnetoresistivity ($\rho_{xx}$) measurements. The angle $\alpha$ is set to zero ($\alpha$=0) when the field is applied parallel to one of the a/b-axis. (b) In-plane angular-dependent magnetoresistivity $\rho_{xx}$, normalized by the minimum value of $\rho_{xx}$ ($\rho_{\mathrm{min}}$), at various temperatures for an applied field of 12 T. (c) Polar plots of the data in (b). (d) Field-dependent ($T$ = 8 K) magnetoresistivity $\rho_{xx}$, normalized by the minimum value of $\rho_{xx}$ ($\rho_{\mathrm{min}}$). (e) Polar plot of the data in (d).}
\label{fig3}
\end{figure*}

Next, we discuss the in-plane anisotropy of the magnetoresistivity in the LiTi$_2$O$_4$ thin films with the Corbino disk geometry [Fig.~\ref{fig3}(a)], which can eliminate the Lorentz force-induced extrinsic anisotropy~\cite{Ref22,Ref23} when the in-plane magnetic field is rotated relative to the crystal axes. $\alpha$ is defined as the azimuthal angle between the magnetic field and the a/b-axis of the lattice, as indicated in Fig.~\ref{fig3}(a). In the normal state [$T$ = 30 K in Fig.~\ref{fig3}(b)], the magnetoresistivity $\rho_{xx}$ is found to be essentially independent of $\alpha$, displaying isotropic behavior. This result is in marked contrast to the twofold rotational symmetry in the magnetoresistivity in the normal state reported for the Hall bar structure in a previous experiment~\cite{Ref15}, suggesting that the previously observed twofold rotational symmetry in the normal state is attributed to an extrinsic effect mainly originated from the Lorentz force as evidenced by the minimum of $\rho_{xx}$ for a field parallel to the current and maximum of $\rho_{xx}$ for a field perpendicular to the current (also see Fig. S5 in the Supplemental Material~\cite{SM}). In the superconducting state [$T$ = 6.5 K in Fig.~\ref{fig3}(b)], we observe a pronounced fourfold modulation of the magnetoresistivity $\rho_{xx}$ [Fig.~\ref{fig3}(c)], which is consistent across multiple samples. In this case, the anisotropic magnetoresistivity $\rho_{xx}$ attains the maximum value when the magnetic field is directed along the [110] and [1$\bar{1}$0] orientations ($\alpha=\pm45^{\circ}$) and becomes minimum when the field is directed along the axes of the lattice ($\alpha=0^{\circ}$ and $90^{\circ}$). Alternatively, the negligible component of twofold symmetry is immersed in the in-plane azimuthal angular-dependent magnetoresistivity $\rho_{xx}$ [$T$ = 8 K in Fig.~\ref{fig3}(b)] and becomes enhanced with increasing field, implying that this twofold symmetry in the superconducting phase transition region mainly originates from the extrinsic contribution induced by the applied magnetic field, such as the magnetic field-induced vortex dynamics or a possible effect of the misalignment of the field with the film plane (see Fig. S6 in the Supplemental Material~\cite{SM}). Considering that the existence of striking fourfold oscillations in magnetoresistivity manifests deep inside the superconducting region and vanishes in the normal state, we can straightforwardly rule out the possibilities of extrinsic contributions, such as the twin structure in the orthorhombic electronic phases~\cite{Ref24,Mao2000} and the Fermi surface of cubic band structure~\cite{Metlushko1997} (see Figs. S7 and S8 and Sec. IV in the Supplemental Material~\cite{SM}) inherent to the LiTi$_2$O$_4$ crystal with respect to the underlying fourfold lattice symmetry shown in Fig.~\ref{fig1}(c), and thus demonstrate that the fourfold rotational symmetry is an intrinsic property of the superconducting phase in LiTi$_2$O$_4$.

\begin{figure*}
\centering
\includegraphics[bb=90 60 470 410,width=11cm,height=9.5cm]{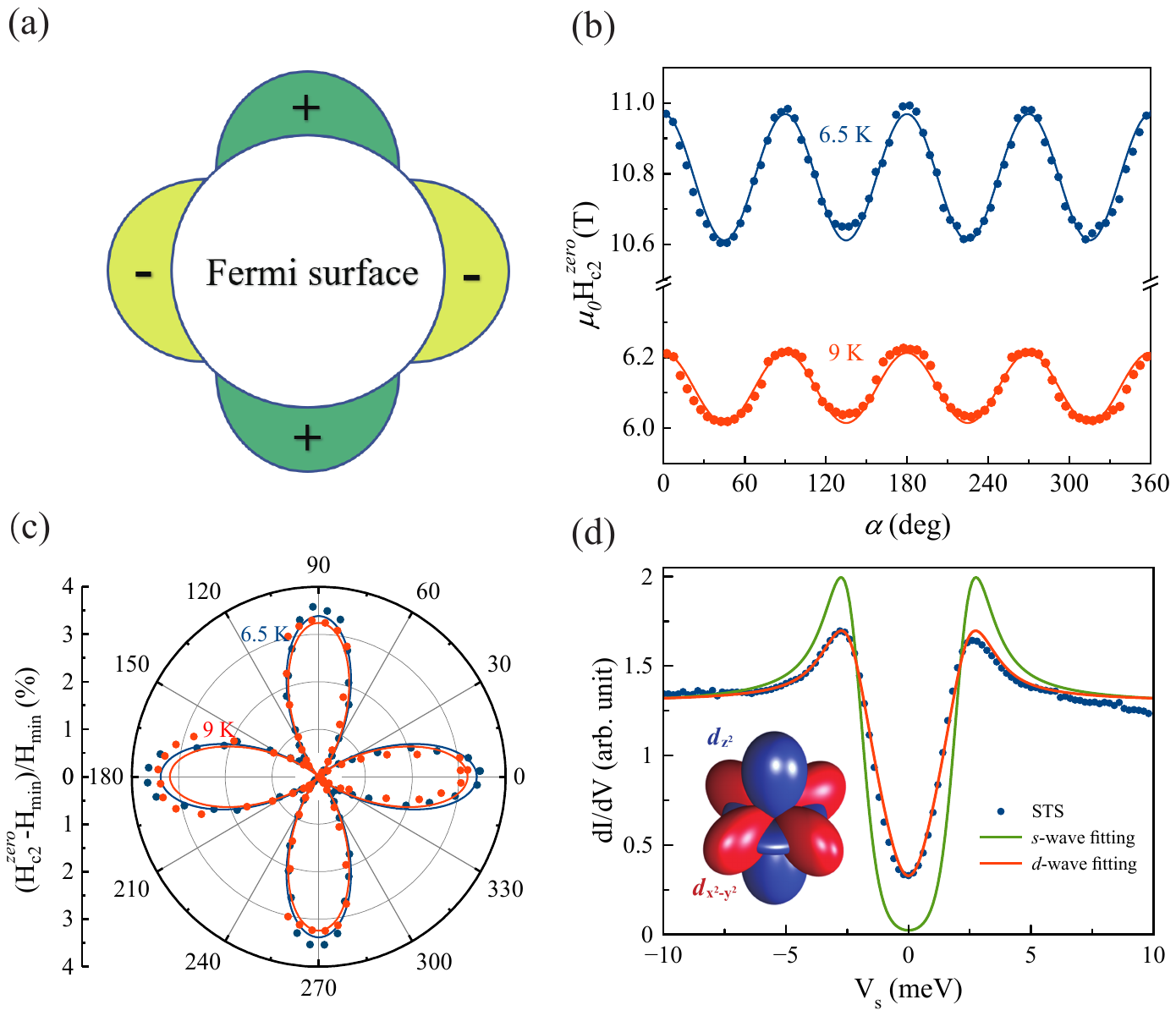}
\caption{$d$-wave pairing signature of LiTi$_2$O$_4$ (001) thin films. (a) Schematic of the in-plane $d_{x^2-y^2}$-wave superconducting gaps on the Fermi surface. (b) In-plane angular-dependent $\mu_0$H$_{c2}^{zero}$ at various temperatures. Here, H$_{c2}^{zero}$ is extracted at the points of zero value of magnetoresistivity $\rho_{xx}$. (c) Polar plots of the data normalized by the minimum value of H$_{c2}^{zero}$ (H$_{\mathrm{min}}$) in (b). (d) High-resolution tunneling conductance spectrum ($dI/dV$, $2\Delta_g$=4.8 meV) on the surface of LiTi$_2$O$_4$ at 4.2 K obtained by using STS. The $s$-wave and $d$-wave with the irreducible representation of $E_g$ fitting curves are also shown in (d), indicating that the superconducting gap symmetry of LiTi$_2$O$_4$ is an unconventional $d$-wave pairing. Here it should be noted that the in-plane $d_{x^2-y^2}$-wave and out-of-plane $d_{z^2}$-wave are degenerate and belong to the two-dimensional irreducible representation of $E_g$ protected by the $O_h$ point group in the cubic LiTi$_2$O$_4$.}
\label{fig4}
\end{figure*}

To further reveal the fourfold symmetric behavior of superconductivity in LiTi$_2$O$_4$ reflecting the superconducting gap structure, we extract the upper critical field $\mu_0$H$_{c2}$ from the $\alpha$-dependent magnetoresistivity $\rho_{xx}$ in the superconducting region. Due to the relatively large value of the upper critical field H$_{c2}$ compared to the strength of the applied magnetic field [see Fig.~\ref{fig2}(d)], we alternatively redefine the points of zero value of magnetoresistivity $\rho_{xx}$, $\mu_0$H$_{c2}^{zero}$, as shown in Fig.~\ref{fig4}(b) and (c). Interestingly, the in-plane $\alpha$-dependent $\mu_0$H$_{c2}^{zero}$ also displays fourfold periodicity, providing additional strong evidence for the fourfold rotational symmetry of the superconductivity in LiTi$_2$O$_4$. In addition, this oscillation of H$_{c2}^{zero}$ has a $\pi$ phase shift compared with that of the magnetoresistivity $\rho_{xx}$ shown in Fig.~\ref{fig3}(c) such that at the $\alpha$ where superconductivity is hardest to suppress, H$_{c2}^{zero}$ is the largest and the magnetoresistivity $\rho_{xx}$ is the lowest [Figs.~\ref{fig3}(c) and~\ref{fig4}(c)], as expected from our intuitions~\cite{Ref25}. Theoretically, the in-plane fourfold symmetric H$_{c2}^{zero}$ has been demonstrated to result from $d$-wave pairing symmetry~\cite{Ref24}. Therefore, the fourfold symmetry in H$_{c2}^{zero}$ inherent to the superconducting Cooper pairs enables us to specify the existence of node directions. Since H$_{c2}^{zero}$ takes maxima for the magnetic field applied parallel to the lattice axes and minima for the directions $45^{\circ}$ from the axes, the superconducting gap leads to a maximum along the lattice axes and a minimum along the [110] and [1$\bar{1}$0] directions ($\alpha=\pm45^{\circ}$), signaling that the in-plane pairing symmetry of the LiTi$_2$O$_4$ most likely belongs to the $d_{x^2-y^2}$-wave associated with the nodes along the directions of H$_{c2}^{zero}$ minima [see Fig.~\ref{fig4}(a)]. In fact, a clear fourfold modulation of H$_{c2}$ in the superconducting state with an applied in-plane magnetic field that reflects the angular positions of nodes of $d_{x^2-y^2}$-wave symmetry has also been observed in cuprates with two-dimensional layered Cu-O planes~\cite{Ref26,Ref27}. On the other hand, the in-plane angular-dependent Hall resistivity $\rho_{xy}(\alpha)$ is also measured on the same footing on the Hall bar geometry (see Fig. S5 in the Supplemental Material~\cite{SM}) to further clarify the $d_{x^2-y^2}$-wave pairing symmetry in nature. Strikingly, the amplitude of $\rho_{xy}(\alpha)$ exhibits fourfold symmetry with extreme values located at ($\alpha=\pm45^{\circ}$), and the phases of the $\rho_{xy}(\alpha)$ change their sign oppositely after the in-plane fourfold rotational operation on rotating magnetic field direction in the superconducting state (see Fig. S5 in the Supplemental Material~\cite{SM}). This finding unambiguously rules out the possibility of the $s$-wave state with anisotropic gap minima and points to the manifestation of the $d$-wave superconducting pairing with nodes along the [110] and [1$\bar{1}$0] directions in LiTi$_2$O$_4$ (also see the detailed theoretical discussions in Sec. VI and VII in the Supplemental Material~\cite{SM}).

\section{Discussions}

Considering that the spinel oxide LiTi$_2$O$_4$ is a three-dimensional superconductor, the cubic symmetry requires that the basis functions of $d_{x^2-y^2}$ and $d_{z^2}$ are degenerate and belong to the two-dimensional irreducible representation of $E_g$ protected by the $O_h$ point group (see the detailed theoretical discussion in Sec. VI in the Supplemental Material~\cite{SM}). This symmetry requirement indicates that the pairing symmetry of LiTi$_2$O$_4$ is indeed an unconventional $d$-wave with the irreducible representation of $E_g$, and the fourfold modulations inherent to the superconducting Cooper pairing amplitude in LiTi$_2$O$_4$ (001) thin films will not only exist in the in-plane directions [Fig.~\ref{fig4}(c)], but also emerge in the out-of-plane orientations. Interestingly, this fourfold modulation is clearly visible for the out-of-plane polar angular-dependent H$_{c2}$ shown in Fig. S11 in the Supplemental Material~\cite{SM}, consistent with the theoretical expectation. Due to the degeneracy of $d_{x^2-y^2}$-wave and $d_{z^2}$-wave pairings, we further expect the existence of in-plane sixfold rotational symmetry in H$_{c2}$ in the (111)-oriented LiTi$_2$O$_4$ thin films. Surprisingly, a pronounced sixfold modulation is found in the in-plane azimuthal angle-dependent H$_{c2}$ when the LiTi$_2$O$_4$ thin films are epitaxially grown on the (111)-oriented MgAl$_2$O$_4$ substrates, as shown in Fig. S12 in the Supplemental Material~\cite{SM}. These complementary electrical transport results unambiguously provide the strong compelling evidences for the $d$-wave superconducting Cooper pair formation in LiTi$_2$O$_4$.

Furthermore, STS measurements are independently carried out to further examine the superconducting gap features. Figure~\ref{fig4}(d) displays a typical tunneling conductance ($dI/dV$) spectrum as a function of a bias voltage ($V_s$) applied at the surface of three-dimensional LiTi$_2$O$_4$ thin films with a fixed temperature of 4.2 K. Remarkably, the tunneling conductance with sharp coherence peaks does not go to zero at the Fermi energy, and a pronounced V-shaped-like behavior rather than the U-shaped behavior is clearly visible, suggestive of the existence of a superconducting gap with nodes having gapless excitations in LiTi$_2$O$_4$. In addition, theoretical models with both $s$-wave and $d$-wave with the two-dimensional irreducible representation of $E_g$ ($d_{x^2-y^2}$+$d_{z^2}$-wave) pairings are quantitatively fitted to the superconducting gap using the Dynes formula~\cite{Ref28,Ref29} (also see the detailed fitting in Sec. VIII in the Supplemental Material~\cite{SM}). Through the theoretical fitting shown in Fig.~\ref{fig4}(d), we find that the theoretical curve of $d$-wave pairing well reproduces the STS spectrum, and the fitted superconducting gap $2\Delta_g$ is 4.8 meV. This result provides further evidence for the $d$-wave pairing superconductivity in LiTi$_2$O$_4$, suggestive of an unconventional superconducting pairing mechanism, such as the anisotropic spin fluctuations induced by charge frustration with the mixed-valence state of Ti$^{3.5+}$ inherent to LiTi$_2$O$_4$.

Therefore, the anisotropic $d$-wave Cooper pair hosted in the spinel oxide LiTi$_2$O$_4$ with three-dimensional superconductivity is an appealing example, followed by the high-$T_c$ cuprates with two-dimensional superconductivity, which not only provides a new platform to clarify the emergence of unconventional superconductivity with a delicate interplay of charge frustration associated with spin fluctuations and Cooper pairs, but also initiate revised theories for the pairing mechanism of unconventional superconductivity.

\section{MATERIALS AND METHODS}

\subsection{Thin film growth and structural characterization}

Spinel LiTi$_2$O$_4$ (001) thin films are epitaxially grown on MgAl$_2$O$_4$ (001) substrates ($4\times4\times0.5$ mm$^3$) by pulsed laser deposition in an ultrahigh vacuum chamber (base pressure of $10^{-8}$ Torr). Prior to growth, the MgAl$_2$O$_4$ substrates are annealed at 800 $^{\circ}$C for 2 hours in air to obtain a smooth surface (Fig. S1 in the Supplemental Material~\cite{SM}). During deposition, a sintered Li$_4$Ti$_5$O$_{12}$ ceramic target (Kurt J. Lesker Company) is used to grow the LiTi$_2$O$_4$ films with a KrF excimer laser (Coherent 102, wavelength: $\lambda$ = 248 nm). A pulse energy of 110 mJ and a repetition rate of 10 Hz are used. The LiTi$_2$O$_4$ films are deposited at 750 $^{\circ}$C in a vacuum chamber to promote growth of the superconducting phase. All the samples are cooled to room temperature at a constant rate of 20 $^{\circ}$C/min in vacuum after deposition. The crystalline quality and epitaxy relationship of LiTi$_2$O$_4$ thin films are examined by four-circle XRD (Bruker D8 Discover, Cu K$\alpha$ radiation, $\lambda$ = 1.5406 \AA) operated in high-resolution mode using a three-bounce symmetric Ge (022) crystal monochromator.

\subsection{Magnetization and electrical transport measurements}

Before electrical transport measurements, the magnetic properties of the LiTi$_2$O$_4$ films are measured using a superconducting quantum interference device (SQUID) magnetometer (MPMS, Quantum Design Inc.). For a measurement of the DC magnetization as a function of temperature, the samples are first cooled to 2 K in zero field, and then, an out-of-plane magnetic field of 50 Oe is applied. The magnetization data are collected during warming from 2 K to 30 K (ZFC process). In the same fixed field, the samples are then cooled to 2 K again, and the magnetization data are recollected during warming from 2 K to 30 K (FC process). The electrical transport measurements are performed using a cryostat (Oxford Instruments TeslatronPT cryostat system). The Hall bar and Corbino~\cite{Ref23} devices are fabricated by ion-beam etching to measure the electrical transport properties, which can be seen clearly in the optical microscopic images shown in Fig. S2 in the Supplemental Material~\cite{SM}. Using a commercially available measurement apparatus, the samples are mounted on a mechanical rotator in a $^4$He cryostat to study the anisotropy of superconductivity. The misalignment of the field with the film plane is estimated to be less than 7$^{\circ}$ as our experimental error.

\subsection{STS measurements}

The features of the superconducting gap in LiTi$_2$O$_4$ thin films are measured using STS cooled by liquid helium. The tunneling conductance ($dI/dV$) spectrum as a function of a bias voltage ($V_s$) applied at the surface of LiTi$_2$O$_4$ thin films with a fixed temperature of 4.2 K is measured with the assistance of a lock-in amplifier, the bias modulation $\Delta U_{\mathrm{rms}}=200 \mu$V ($f$ = 983 Hz) is applied, and the tip-sample distance is set by $U$ = 10 mV and $I$ = 100 pA. The single-band Dynes formulas~\cite{Ref28,Ref29} with $s$-wave and $d$-wave pairing formations are used for fitting the tunneling conductance ($dI/dV$) shown in Fig.~\ref{fig4}(d) (also see the details in Sec. VIII of the Supplemental Material~\cite{SM}).

\section*{ACKNOWLEDGMENTS}

This work is supported by the National Natural Science Foundation of China (Grant Nos. 11927807 and 12027805) and Shanghai Science and Technology Committee (Grant Nos. 19ZR1402600 and 20DZ1100604).

\notag\

The authors declare that they have no competing interests.

\notag\

W.L. conceived the project and designed the experiments. H.X. grew the samples. L.W. and W.P. performed the XRD measurements. L.W. performed the electrical transport and magnetization measurements. H.X. and Z.A. did the sample nano-fabrications. Z.W. and C.G. performed the STS measurements. W.L. wrote the paper. All authors discussed the results and gave approval to the final version of the manuscript.

\notag\

\noindent $^*$These authors contributed equally to this work.\\
\noindent $^{\dag}$To whom correspondence should be addressed. E-mail: clgao@fudan.edu.cn\\
\noindent $^{\ddag}$To whom correspondence should be addressed. E-mail: anzhenghua@fudan.edu.cn\\
\noindent $^{\S}$To whom correspondence should be addressed. E-mail: w$\_$li@fudan.edu.cn


\begin{thebibliography}{99}

\bibitem{Ref1} M. Sigrist and K. Ueda, {\it Phenomenological theory of unconventional superconductivity}, Rev. Mod. Phys. \textbf{63}, 239 (1991).

\bibitem{Ref2} C. C. Tsuei and J. R. Kirtley, {\it Pairing symmetry in cuprate superconductors}, Rev. Mod. Phys. \textbf{72}, 969 (2000).

\bibitem{Ref3} R. N. Michael, {\it The challenge of unconventional superconductivity}, Science \textbf{332}, 196 (2011).

\bibitem{Ref4} G. R. Stewart, {\it Unconventional superconductivity}, Adv. Phys. \textbf{66}, 75 (2017).

\bibitem{Ref5} J. Bardeen, L. N. Cooper, and J. R. Schrieffer, {\it Theory of superconductivity}, Phys. Rev. \textbf{108}, 1175 (1957).

\bibitem{Ref6} Z.-X. Shen, D. S. Dessau, B. O. Wells, D. M. King, W. E. Spicer, A. J. Arko, D. Marshall, L. W. Lombardo, A. Kapitulnik, P. Dickinson, S. Doniach, J. DiCarlo, T. Loeser, and C. H. Park, {\it Anomalously large gap anisotropy in the a-b plane of Bi$_2$Sr$_2$CaCu$_2$O$_{8+\delta}$}, Phys. Rev. Lett. \textbf{70}, 1553 (1993).

\bibitem{Ref7} C. C. Tsuei, J. R. Kirtley, C. C. Chi, L. S. Yu-Jahnes, A. Gupta, T. Shaw, J. Z. Sun, and M. B. Ketchen, {\it Pairing symmetry and flux quantization in a tricrystal superconducting ring of YBa$_2$Cu$_3$O$_{7-\delta}$}, Phys. Rev. Lett. \textbf{73}, 593 (1994).

\bibitem{Ref8} J. Wu, A. T. Bollinger, X. He, and I. Bo\v{z}ovi\'{c}, {\it Spontaneous breaking of rotational symmetry in copper oxide superconductors}, Nature \textbf{547}, 432 (2017).

\bibitem{DHLee2013} J. C. S. Davis and D.-H. Lee, {\it Concepts relating magnetic interactions, intertwined electronic orders, and strongly correlated superconductivity}, Proc. Natl. Acad. Sci. U. S. A. \textbf{110}, 17623 (2013).

\bibitem{Ref9} D. C. Johnston, H. Prakash, W. H. Zachariasen, and R. Viswanathan, {\it High temperature superconductivity in the Li-Ti-O ternary system}, Mater. Res. Bull. \textbf{8}, 777 (1973).

\bibitem{Ref10} R. W. McCallum, D. C. Johnston, C. A. Luengo, and M. B. Maple, {\it Superconducting and normal state properties of Li$_{1+x}$Ti$_{2-x}$O$_4$ spinel compounds. II. Low-temperature heat capacity}, J. Low Temp. Phys. \textbf{25}, 177 (1976).

\bibitem{SunCP2004} C. P. Sun, J.-Y. Lin, S. Mollah, P. L. Ho, H. D. Yang, F. C. Hsu, Y. C. Liao, and M. K. Wu, {\it Magnetic field dependence of low-temperature specific heat of the spinel oxide superconductor LiTi$_2$O$_4$}, Phys. Rev. B \textbf{70}, 054519 (2004).

\bibitem{Ref11} O. Durmeyer, J. P. Kappler, E. Beaurepaire, J. M. Heintz, and M. Drillon, {\it Ti K XANES in superconducting LiTi$_2$O$_4$ and related compounds}, J. Phys.: Condens. Matter \textbf{2}, 6127 (1990).

\bibitem{Ref12} C. L. Chen, C. L. Dong, K. Asokan, J. L. Chen, Y. S. Liu, J.-H. Guo, W. L. Yang, Y. Y. Chen, F. C. Hsu, C. L. Chang, and M. K. Wu, {\it Role of $3d$ electrons in the rapid suppression of superconductivity in the dilute V doped spinel superconductor LiTi$_2$O$_4$}, Supercon. Sci. Technol. \textbf{24}, 115007 (2011).

\bibitem{Ref13} D. P. Tunstall, J. R. M. Todd, S. Arumugam, G. Dai, M. Dalton, and P. P. Edwards, {\it Titanium nuclear magnetic resonance in metallic superconducting lithium titanate and its lithium-substituted derivatives Li$_{1+x}$Ti$_{2-x}$O$_4$ ($0<x<0.10$)}, Phys. Rev. B \textbf{50}, 16541 (1994).

\bibitem{Ref14} D. C. Johnston, {\it Superconducting and normal state properties of Li$_{1+x}$Ti$_{2-x}$O$_4$ spinel compounds. I. Preparation, crystallography, superconducting properties, electrical resistivity, dielectric behavior, and magnetic susceptibility}, J. Low Temp. Phys. \textbf{25}, 145 (1976).

\bibitem{Ref15} K. Jin, G. He, X. Zhang, S. Maruyama, S. Yasui, R. Suchoski, J. Shin, Y. Jiang, H. S. Yu, J. Yuan, L. Shan, F. V. Kusmartsev, R. L. Greene, and I. Takeuchi, {\it Anomalous magnetoresistance in the spinel superconductor LiTi$_2$O$_4$}, Nat. Commun. \textbf{6}, 7183 (2015).

\bibitem{Ref16}	Y. Okada, Y. Ando, R. Shimizu, E. Minamitani, S. Shiraki, S. Watanabe, and T. Hitosugi, {\it Scanning tunnelling spectroscopy of superconductivity on surfaces of LiTi$_2$O$_4$(111) thin films}, Nat. Commun. \textbf{8}, 15975 (2017).

\bibitem{SM} See Supplemental Material for more discussions on the extra experimental data, the first-principles calculatons, and the relevant theoretical fittings. 

\bibitem{Ref17} M. Tinkham, {\it Introduction to Superconductivity}, 2nd edn (McGraw-Hill, New York, 1996).

\bibitem{Ref18} D. Jiang, T. Yuan, Y. Wu, X. Wei, G. Mu, Z. An, and W. Li, {\it Strong in-plane magnetic field-induced reemergent superconductivity in the van der Waals heterointerface of NbSe$_2$ and CrCl$_3$}, ACS Appl. Mater. Interfaces \textbf{12}, 49252 (2020).

\bibitem{Ref19} N. R. Werthamer, E. Helfand, and P. C. Hohenberg, {\it Temperature and purity dependence of the superconducting critical Field, H$_{c2}$. III. Electron spin and spin-orbit effects}, Phys. Rev. \textbf{147}, 295 (1966).

\bibitem{Ref20} Z. Wei, G. He, W. Hu, Z. Feng, X. Wei, C. Y. Ho, Q. Li, J. Yuan, C. Xi, Z. Wang, Q. Chen, B. Zhu, F. Zhou, X. Dong, L. Pi, A. Kusmartseva, F. V. Kusmartsev, Z. Zhao, and K. Jin, {\it Anomalies of upper critical field in the spinel superconductor LiTi$_2$O$_{4-\delta}$}, Phys. Rev. B \textbf{100}, 184509 (2019).

\bibitem{Ref21} R. V. Chopdekar, F. J. Wong, Y. Takamura, E. Arenholz, and Y. Suzuki, {\it Growth and characterization of superconducting spinel oxide LiTi$_2$O$_4$ thin films}, Physica C \textbf{469}, 1885 (2009).

\bibitem{Ref22} Y. Paltiel, E. Zeldov, Y. Myasoedov, M. L. Rappaport, G. Jung, S. Bhattacharya, M. J. Higgins, Z. L. Xiao, E. Y. Andrei, P. L. Gammel, and D. J. Bishop, {\it Instabilities and disorder-driven first-order transition of the vortex lattice}, Phys. Rev. Lett. \textbf{85}, 3712 (2000).

\bibitem{Ref23} J. Li, P. J. Pereira, J. Yuan, Y.-Y. Lv, M.-P. Jiang, D. Lu, Z.-Q. Lin, Y.-J. Liu, J.-F. Wang, L. Li, X. Ke, G. Van Tendeloo, M.-Y. Li, H.-L. Feng, T. Hatano, H.-B. Wang, P.-H. Wu, K. Yamaura, E. Takayama-Muromachi, J. Vanacken, L. F. Chibotaru, and V. V. Moshchalkov, {\it Nematic superconducting state in iron pnictide superconductors}, Nat. Commun. \textbf{8}, 1880 (2017).

\bibitem{Ref24} K. Takanaka and K. Kuboya, {\it Anisotropy of upper critical field and pairing symmetry}, Phys. Rev. Lett. \textbf{75}, 323 (1995).

\bibitem{Mao2000} Z. Q. Mao, Y. Maeno, S. NishiZaki, T. Akima, and T. Ishiguro, {\it In-plane anisotropy of upper critical field in Sr$_2$RuO$_4$}, Phys. Rev. Lett. \textbf{84}, 991 (2000).

\bibitem{Metlushko1997} V. Metlushko, U. Welp, A. Koshelev, I. Aranson, G. W. Crabtree, and P. C. Canfield, {\it Anisotropic upper critical field of LuNi$_2$B$_2$C}, Phys. Rev. Lett. \textbf{79}, 1738 (1997).

\bibitem{Ref25} A. Hamill, B. Heischmidt, E. Sohn, D. Shaffer, K.-T. Tsai, X. Zhang, X. Xi, A. Suslov, H. Berger, L. Forr\'{o}, F. J. Burnell, J. Shan, K. F. Mak, R. M. Fernandes, K. Wang, and V. S. Pribiag, {\it Two-fold symmetric superconductivity in few-layer NbSe$_2$}, Nat. Phys. \textbf{17}, 949 (2021).

\bibitem{Ref26} T. Hanaguri, T. Fukase, Y. Koike, I. Tanaka, and H. Kojima, {\it Anisotropy of upper critical field in the (110)$_t$ and (001)$_t$ planes for single-crystal La$_{1.86}$Sr$_{0.14}$CuO$_4$}, Physica B \textbf{165-166}, 1449 (1990).

\bibitem{Ref27} Y. Koike, T. Takabayashi, T. Noji, T. Nishizaki, and N. Kobayashi, {\it Fourfold symmetry in the $ab$ plane of the upper critical field for single-crystal Pb$_2$Sr$_2$Y$_{0.62}$Ca$_{0.38}$Cu$_3$O$_8$: Evidence for $d_{x^2-y^2}$ pairing in a high-$T_c$ superconductor}, Phys. Rev. B \textbf{54}, R776 (1996).

\bibitem{Ref28} R. C. Dynes, V. Narayanamurti, and J. P. Garno, {\it Direct measurement of quasiparticle-lifetime broadening in a strong-coupled superconductor}, Phys. Rev. Lett. \textbf{41}, 1509 (1978).

\bibitem{Ref29} T. Machida, Y. Kohsaka, and T. Hanaguri, {\it A scanning tunneling microscope for spectroscopic imaging below 90 mK in magnetic fields up to 17.5 T}, Rev. Sci. Instrum. \textbf{89}, 093707 (2018).


\end{thebibliography}
\end{document}